\documentclass[twocolumn,prl,amsmath,amssymb,showpacs,superscriptaddress]{revtex4-1}
\usepackage{epsf}      
\usepackage{graphicx}
\usepackage{color}
\usepackage{gensymb}
\usepackage{sidecap}

\begin{document}
\title {Tuning ferromagnetism and spin state in La$_{(1-x)}$$A_x$CoO$_3$ ($A=$ Sr, Ca) nanoparticles}
\author{Ravi Prakash}
\email{These authors contributed equally to this work}
\affiliation{Department of Physics, Indian Institute of Technology Delhi, Hauz Khas, New Delhi-110016, India}
\author{Rishabh Shukla}
\email{These authors contributed equally to this work}
\affiliation{Department of Physics, Indian Institute of Technology Delhi, Hauz Khas, New Delhi-110016, India}
\author{Priyanka}
\affiliation{Department of Physics, Indian Institute of Technology Delhi, Hauz Khas, New Delhi-110016, India}
\author{Anita Dhaka}
\affiliation{Department of Physics, Sri Aurobindo College, University of Delhi, Malviya Nagar, New Delhi-110017, India}
\author{R. S. Dhaka}
\email{rsdhaka@physics.iitd.ac.in}
\affiliation{Department of Physics, Indian Institute of Technology Delhi, Hauz Khas, New Delhi-110016, India}

\date{\today}                                         

\begin{abstract}

We use the x-ray diffraction, magnetic susceptibility, isothermal magnetization, and photoelectron spectroscopy to study the structural, magnetic and electronic properties of La$_{(1-x)}$$A_x$CoO$_3$ ($A=$ Sr, Ca; $x=$ 0 -- 0.2) nanoparticles. The Rietveld refinements of room temperature powder x-ray diffraction data confirm the single phase and the rhombohedral crystal structure with R$\bar{3}$C space group. We find drastic changes in the magnetic properties and spin-states with Sr/Ca substitution (hole doping). For $x=$ 0 sample, the magnetic measurements show a ferromagnetic transition at T$_{\rm C}$$\approx$85 K, which shifted significantly to higher temperatures with hole doping; simultaneously a significant increase in the spontaneous magnetic moment has been observed. Whereas, the coercive field H$_{\rm C}$ values are 7, 4.4 and 13.2~kOe for $x=$ 0, 0.2 (Sr), and 0.2(Ca) samples. Furthermore, the FC magnetization shows a ferromagnetic Brillouin function like behavior at low temperatures for Ca samples. We demonstrate that the Sr/Ca substitution increases the population of IS (Co$^{3+}$) and LS (Co$^{4+}$) states and tune the ferromagnetism in nanoparticles via double-exchange interactions between Co$^{3+}$-- Co$^{4+}$. Our results suggest an important role of hole carriers and nano-size effect in controlling the spin-state and magnetism in La$_{(1-x)}$$A_x$CoO$_3$ nanoparticles.\\


\end{abstract}

\maketitle

\section{\noindent ~Introduction}

The perovskite cobaltites La$_{1-x}$A$_x$CoO$_3$ (A = alkaline-earth ions) have attracted much attention due to their remarkable magnetic and transport properties such as colossal magnetoresistance (CMR), large thermoelectric effect, spin-state and metal-insulator transitions \cite{Briceno,Kriener,Wu,Aarbogh}. It is believed that the presence of strong electron-electron correlations and interplay between charge, orbital and spin degrees of freedom are playing crucial role in these unique properties. In the LaCoO$_3$ (a rhombohedrally distorted perovskite), the Co$^{3+}$ (3d$^6$) ion is surrounded by six octahedrally-coordinated O$^{2-}$ ions, and due to crystal field splitting the fivefold degenerate 3d orbitals split into the doubly degenerate e$_g$ and the triply degenerate t$_{2g}$ orbitals in the higher and lower energy levels, respectively. It shows a nonmagnetic insulating ground state (spin gap $\approx$30~meV and charge gap $\approx$0.1~eV) with Co$^{3+}$ ions in a low spin (LS; t$^6_{2g}$e$^0_g$ (S = 0)) configuration below 90~K \cite{Korotin96, Yamaguchi96}. By increasing the temperature, a paramagnetic (PM) insulating state develops above 90~K, where a spin-state transition occurs from LS to intermediate-spin (IS; t$^5_{2g}$e$^1_g$ (S = 1)) or high-spin (HS; t$^4_{2g}$e$^2_g$ (S = 2)) or mixed states \cite{Lengsdorf}. Note that there is a subtle competition between intra-atomic Hund's coupling (H$_{ex}$) and the crystal field splitting ($\Delta_{\rm CF}$) energies. The Co$^{3+}$ ions take the HS state when the H$_{ex}$ is larger than the $\Delta_{\rm CF}$, while they take the LS state in the opposite case \cite{Korotin96}. On the other hand, the IS state has been proposed to be energetically close to the LS state when the hybridization between the Co e$_g$ and O 2$p$ orbitals is taken into consideration \cite{Korotin96}, which also found in SrCoO$_3$ using x-ray absorption spectroscopy and theoretical calculations \cite{Potze95} .

	Interestingly, the partial substitution of La$^{3+}$ ions by divalent alkaline-earth ions ($A$) is playing crucial role in tuning the magnetic and transport properties of LaCoO$_3$ \cite{Caciuffo,JiangPRB09,FitaPRB05}. This is due to the addition of holes into the system, which formally creates Co$^{4+}$ ions, and the structural changes because of the different ionic radii of $A$ element. The magnetic phase diagram of La$_{1-x}$Sr$_x$CoO$_3$ indicates the evolution from nonmagnetic insulating spin-glass ($x\le$0.18) to ferromagnetic metallic cluster-glass behavior (0.18$\le x \le$ 0.5), which are ascribed to spin states transitions \cite{Wu}. The spin-glass state is achieved due to the frustration between the antiferromagnetic (AFM) superexchange (Co$^{3+}$--Co$^{3+}$ and Co$^{4+}$--Co$^{4+}$) and the ferromagnetic double exchange (Co$^{3+}$--Co$^{4+}$) interactions. In case of $A=$ Ca, the magnetic and insulator-metal transitions occur at $x \approx$ 0.05 and $\ge$0.2, respectively \cite{Kriener}. It has been shown that hole doping by substitution of larger Sr$^{2+}$ ions (1.31~$\rm \AA$) decreases the crystal field splitting, which may stabilizes the IS state; however, it would be more interesting when the similar size Ca$^{2+}$ (1.18~$\rm \AA$) ions are being substituted at La$^{3+}$ (1.216~$\rm \AA$) site \cite{Shannon76}. The structural and magnetic properties of these materials are highly sensitive to the size and substitution level ($x$) of $A$ element as well as the sample preparation method. There have been many reports on the bulk samples prepared by solid-state reaction \cite{AarboghPRB06,SamalJPCM11,SamalJAP09,SamalJAP12,MannaJAP13,DevendraJPCM13}; however, studies of LaCoO$_3$ nanoparticles (prepared by sol-gel method) are still very limited \cite{ZhouPRB07, FitaPRB08, ZhouJPCC09, DurandJPCM15}. 
	
	In recent years, there is an enormous interest in the LaCoO$_3$  nanoparticles mainly due to their peculiar properties such as ferromagnetism, spin-state and metal-insulator transitions (MIT) with temperature. It is reported that, in contrast to a nonmagnetic insulating ground state of bulk LaCoO$_3$, the nanoparticle and thin-film samples exhibit a ferromagnetic (FM) ordering below T$_{\rm C}$ $\approx$85~K \cite{ZhouPRB07, FitaPRB08, ZhouJPCC09, DurandJPCM15,FuchsPRB09}. The infrared (IR) absorption spectra give evidence for a stabilization of higher spin state and a reduced Jahn-Teller (JT) distortion in the nanoparticles with respect to the bulk LaCoO$_3$, \cite{ZhouPRB07} which is consistent with the results on the strained films \cite{FuchsPRB07}. Jiang {\it et al.} used extended x-ray absorption fine structure to study the La$_{(1-x)}$Sr$_x$CoO$_3$ systems, and found no clear evidence for a JT distortion, which does not favor IS state as it requires a single electron in the Co e$_g$ state and such configuration is expected to be JT active \cite{JiangPRB09}. However, a theoretical study by Pandey {\it et al.} indicates that spin-orbit coupling also splits the e$_g$ degeneracy without a JT distortion \cite{PandeyPRB08}. If such a splitting is sufficient, then an IS state might be achieved without a JT distortion \cite{PandeyPRB08}. It has been shown that monoclinic/rhombohedral distortion affects the electronic structure and orbital ordering-driven ferromagnetism in LaCoO$_3$ \cite{WangJAP10, LeeJAP13, FuchsPRB09}. Yan {\it et al.} observed the ferromagnetism in LaCoO$_3$, which demonstrated to be due to the coupling between the surface Co$^{+3}$ ions \cite{YanPRB04}. In La$_{0.5}$$A_{0.5}$CoO$_3$ ($A=$Ca, Ba, Sr) nanoparticles, it is reported that Co$^{3+}$ ions are in the IS state, but Co$^{4+}$ ions prefer a mixture of IS and HS states \cite{RoyAPL08JAC11}. The magnetic properties of La$_{1-x}$A$_x$CoO$_3$ depend on the spin state of both Co$^{3+}$ and Co$^{4+}$ ions. A size-dependent insulator to metal transition and glassy magnetic behavior have been found in La$_{0.5}$Sr$_{0.5}$CoO$_3$ at nanoscale \cite{RoyAPL08JAC11, LiuAPL09}. Here, an enhancement of the unit cell volume, surface to volume ratio, and strain are considered to play very important role for the appearance of surface FM ordering. However, the nature/origin of the FM state and spin-state transition has been the subject of debate and largely unexplored in hole doped LaCoO$_3$ nanoparticles \cite{FitaJAP10,RoyAPL08JAC11}.  
	
	In this paper, we present detailed investigation of La$_{(1-x)}$$A_x$CoO$_3$ ($A=$ Sr, Ca; $x=$ 0 -- 0.2) nanoparticles using x-ray diffraction, magnetic susceptibility, isothermal magnetization, and x-ray photoelectron spectroscopy (XPS) measurements. We find significant changes in the magnetic behavior with hole doping due to nano-size effect and enhancement in the population of IS (Co$^{3+}$) and LS (Co$^{4+}$) states with Sr/Ca substitution. Our study reveals an important role of hole carriers and nano-scale size in controlling the spin-state and magnetism in hole doped LaCoO$_3$ nanoparticles.

\section{\noindent ~Experimental Details}

The nanoparticles of La$_{(1-x)}$$A_x$CoO$_3$ ($A=$ Sr, Ca; $x=$0 -- 0.2) are synthesized using a sol-gel method in which citric acid was used as gelling agent. The starting materials La(NO$_3$)$_3$ 6H$_2$O (99.99\%), Co(NO$_3$)$_2$ 6H$_2$O (99.999\%), Sr(NO$_3$)$_2$ (99\%) and Ca(NO$_3$)$_2$ (99\%) from Sigma Aldrich were used without further purification. The stoichiometric amounts of these chemicals were dissolved in 2-propyl alcohol to make a clear solution. The solution is then stirred at room temperature (RT) for 24 hrs and evaporated from the open beaker with stirring at 70$\degree$C until the dark purple gel was formed. Evaporation was prolonged at RT for 4 hours and then at 70$\degree$C for 12 hours. After the evaporation, a powder type layer was settled at the bottom of the beaker, which was then dried at 100$\degree$C for 24 hours with heating rate 1$\degree$C/min and finally, sintered these powder samples at 900$\degree$C in air for 6 hrs. We use energy dispersive x-ray spectroscopy (EDX) and powder x-ray diffraction (XRD) with Cu K$\alpha$ ($\lambda$ = 1.5406 $\rm\AA$) radiation to check the overall stoichiometry and the crystal structure/phase purity of all the prepared nanoparticle samples, respectively. We analyzed the XRD data by Rietveld refinement using FULLPROF package and the background was fitted using linear interpolation between the data points. The magnetic measurements are performed using VSM mode in a physical property measurement system (PPMS) from Quantum Design, USA. A commercial electron energy analyzer (PHOIBOS 150 from Specs GmbH, Germany) and a nonmonochromatic MgK$_{\alpha}$ x-ray source (h$\nu=$ 1253.6~eV) have been used to perform the photoemission measurements at room temperature with the base pressure in 10$^{-10}$ mbar range. We have analyzed the core-level spectra \cite{DhakaPRL, DhakaPRB08, DhakaPRB11} after subtracting the inelastic background (by Tougaard method) and MgK$_{\alpha_{3,4}}$ x-ray satellites. 
\begin{figure}[h]
\includegraphics[scale=0.4]{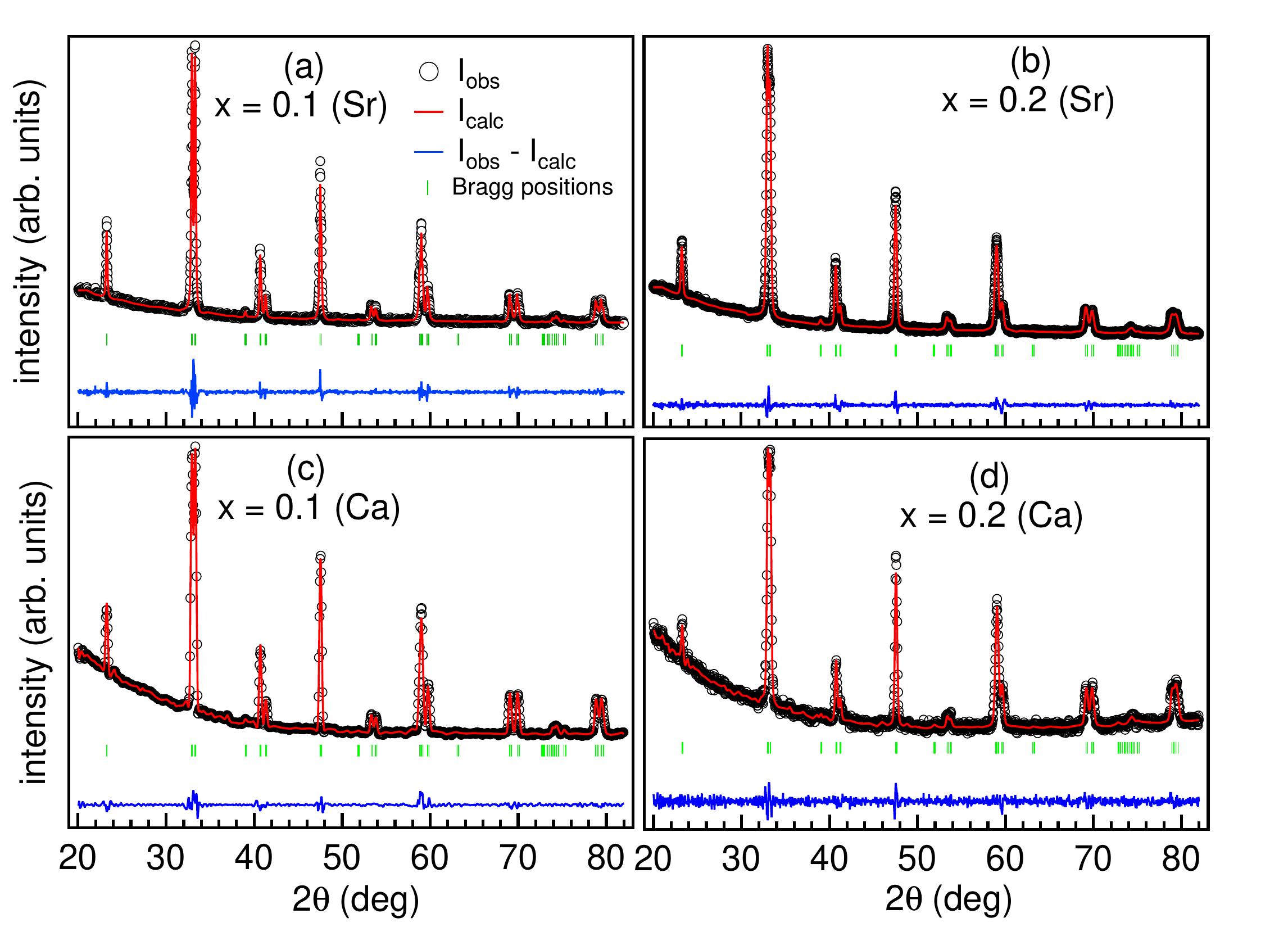}
\caption{Powder XRD data (open circles) and Rietveld refinement (red line) of La$_{(1-x)}$$A_x$CoO$_3$ ($A=$ Sr, Ca) nanoparticles with difference profile (blue line) and Bragg peak positions.}
\label{fig1}
\end{figure}

\section{\noindent ~Results and Discussion}

The room temperature powder x-ray diffraction patterns of La$_{(1-x)}$$A_x$CoO$_3$ ($A=$ Sr, Ca) nanoparticle samples are shown in Fig.~1. 
\begin{figure}[h]
\includegraphics[width=3.4in]{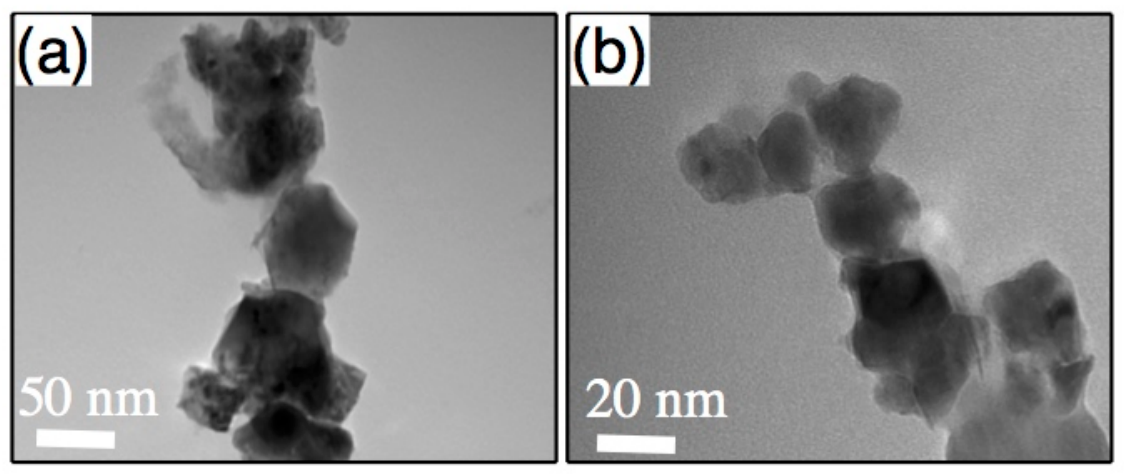}
\caption{Transmission electron micrograph (TEM) of (a) La$_{0.95}$Sr$_{0.05}$CoO$_3$ and (b) La$_{0.9}$Ca$_{0.1}$CoO$_3$ nanoparticles.}
\label{fig2}
\end{figure}
The Rietveld refinements indicate the single phase and confirm the rhombohedral crystal structure with R$\bar{3}$C space group for all the samples (as shown in Fig.~1) and the refined unit cell parameters (not shown) are in good agreement with the reported values \cite{ZhouPRB07,ZhouJPCC09}. The Rietveld refinement and XPS analysis infer the desired $x$- values of Sr/Ca and stoichiometry of oxygen content (3.00$\pm$0.03). Note that Troyanchuk {\it et al.}, reported the change in the crystal structure from rhombohedral (R$\bar{3}$c) to cubic (Pm$\bar{3}$m) with oxygen deficit ($\delta =$ 0.14) \cite{TroyanchukJAP13}. However, our Rietveld analysis of all the samples clearly show the rhombohedral (R$\bar{3}$c) structure, which further confirm that there is no measurable deviation in the oxygen stoichiometry. The average crystallite size has been estimated from the XRD data using Scherer's formula $D=$0.89$\lambda$/$B$ cos $\theta$ (where $\lambda$ is the x-ray wavelength and $B$ is the full width at half maximum (FWHM) of XRD peaks), which comes out to be around 50~nm. The transmission electron micrographs (TEM) show the particle size of 50--100~nm diameter, as shown in Figs.~2 (a, b), which found to be in the similar range of the calculated average crystallite size from XRD data. The samples are also characterized by scanning electron microscopy (SEM), which confirms uniform distribution of the particles (not shown). 

In Figs.~3 (a, b), we present the magnetic behavior of LaCoO$_3$ ($x=$ 0) nanoparticles (prepared by sol-gel method) and bulk (prepared by solid-state reaction method) samples by comparing the temperature dependence of the zero field cooled (ZFC) and field cooled (FC) magnetization (M--T) and isothermal magnetization hysteresis (M--H) data measured at H = 1~kOe and T = 5~K, respectively. The M--T data show a clear ferromagnetic transition at T$_{\rm C}$$\approx$80~K for both the samples \cite{YanPRB04, ZhouPRB07}; however, the magnetic behavior below T$_{\rm C}$ is significantly different. For example, below T$_{\rm C}$ a significant decrease in ZFC magnetization is observed till 22~K, which then increases sharply down to 5~K in the bulk sample. On the other hand, for nanoparticle sample the ZFC magnetization increases till 40~K and then decreases with decreasing the temperature. The ZFC curve shows a broad cusp at about 40~K, which is usually believed to be the characteristics of a glasslike behavior \cite{Wu}. The FC magnetization increases linearly with decreasing the temperature below T$_{\rm C}$ for nanoparticle sample and reached about 84~emu/mol at 5~K; however, for bulk sample it increases slowly till 20~K and then there is a fast increase to 18~emu/mol at 5~K. A large bifurcation in ZFC and FC curves below 50~K indicates the presence of short-range ferromagnetic interactions. Above T$_C$, the nanoparticles exhibit a paramagnetic state, where the magnetic susceptibility obeys the Curie-Weiss law (discussed later). The isothermal magnetization (M--H) data measured at 5~K (Fig.~3b) show no saturation even up to 7~tesla for both the samples, which indicate the canted AFM behavior \cite{Vinod14}. As mentioned before, the superexchange interaction between Co$^{3+}$--Co$^{3+}$ ions gives the AFM contribution, which can be enhanced due to additional Co$^{4+}$--Co$^{4+}$ in hole doped LaCoO$_3$ \cite{ZhouJAP13}. Moreover, the coercivity and remanence magnetization increases significantly in nanoparticles (H$_{\rm C}=$ 7.0~kOe, M$_{\rm r}=$ 85~emu/mole) compare to bulk (H$_{\rm C}=$  1.35~kOe, M$_{\rm r}=$ 11~emu/mole) sample at 5~K. The value of coercive field is comparable with the reported values of the ground single crystals and strained films \cite{YanPRB04, FuchsPRB07}. Our magnetization results [Figs.~3(a, b)] for both bulk and nanoparticle samples ($x=$ 0) are in good agreement with refs.~\onlinecite{Vinod14} and \onlinecite{ZhouPRB07}, respectively. These measurements again confirm the oxygen stoichiometry close to 3 in the present case because the deviation from 3 will decrease the magnetization, as reported in ref.~\cite{HaggertyJPCM04}.

\begin{figure}[h]
\includegraphics[width=3.4in]{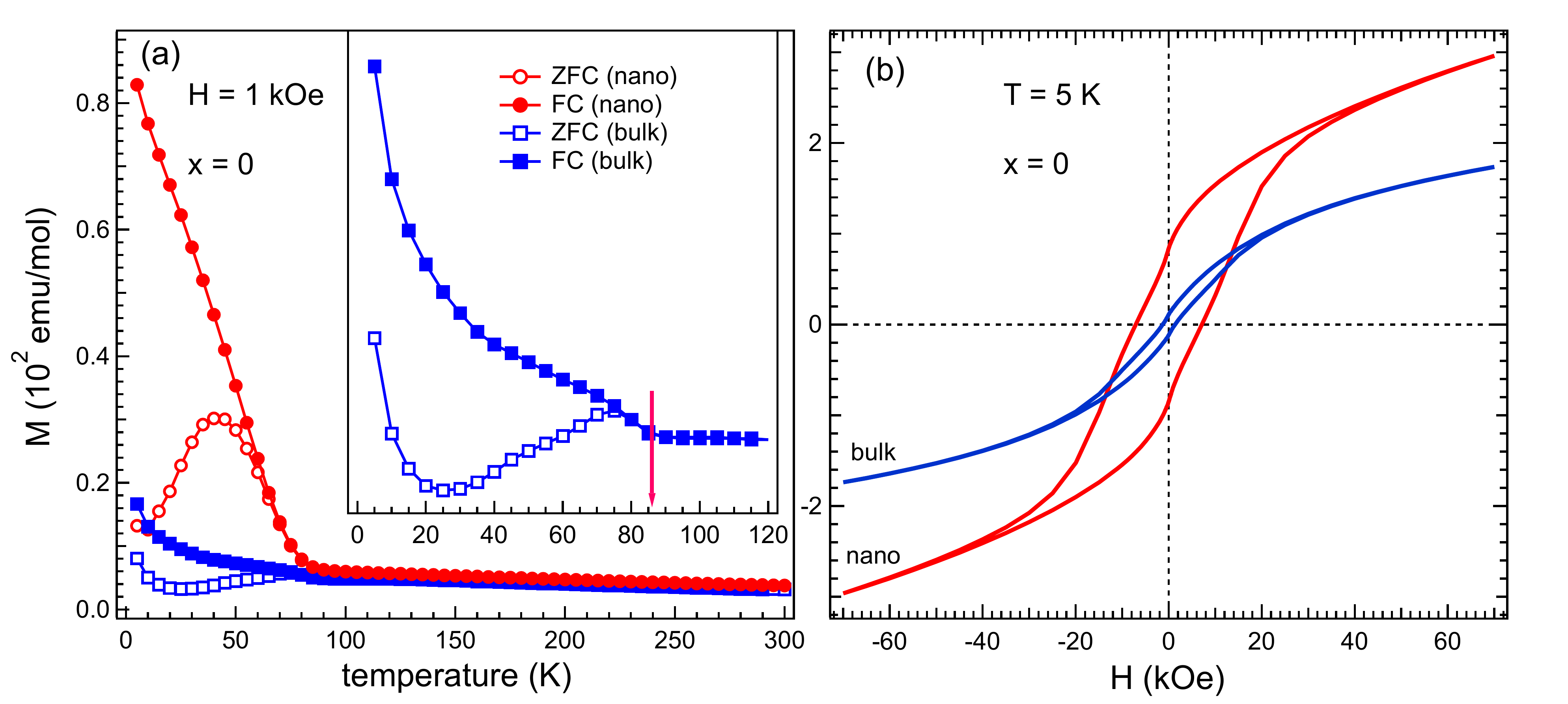}
\caption{The magnetic behavior of LaCoO$_3$ (bulk and nanoparticles). (a) The FC (solid symbol) and ZFC (open symbol) magnetic susceptibility data, measured at 1~KOe. (b) Isothermal M-H loops, measured at T = 5 K.}
\label{fig3}
\end{figure}

\begin{figure}[h]
\includegraphics[width=3.55in]{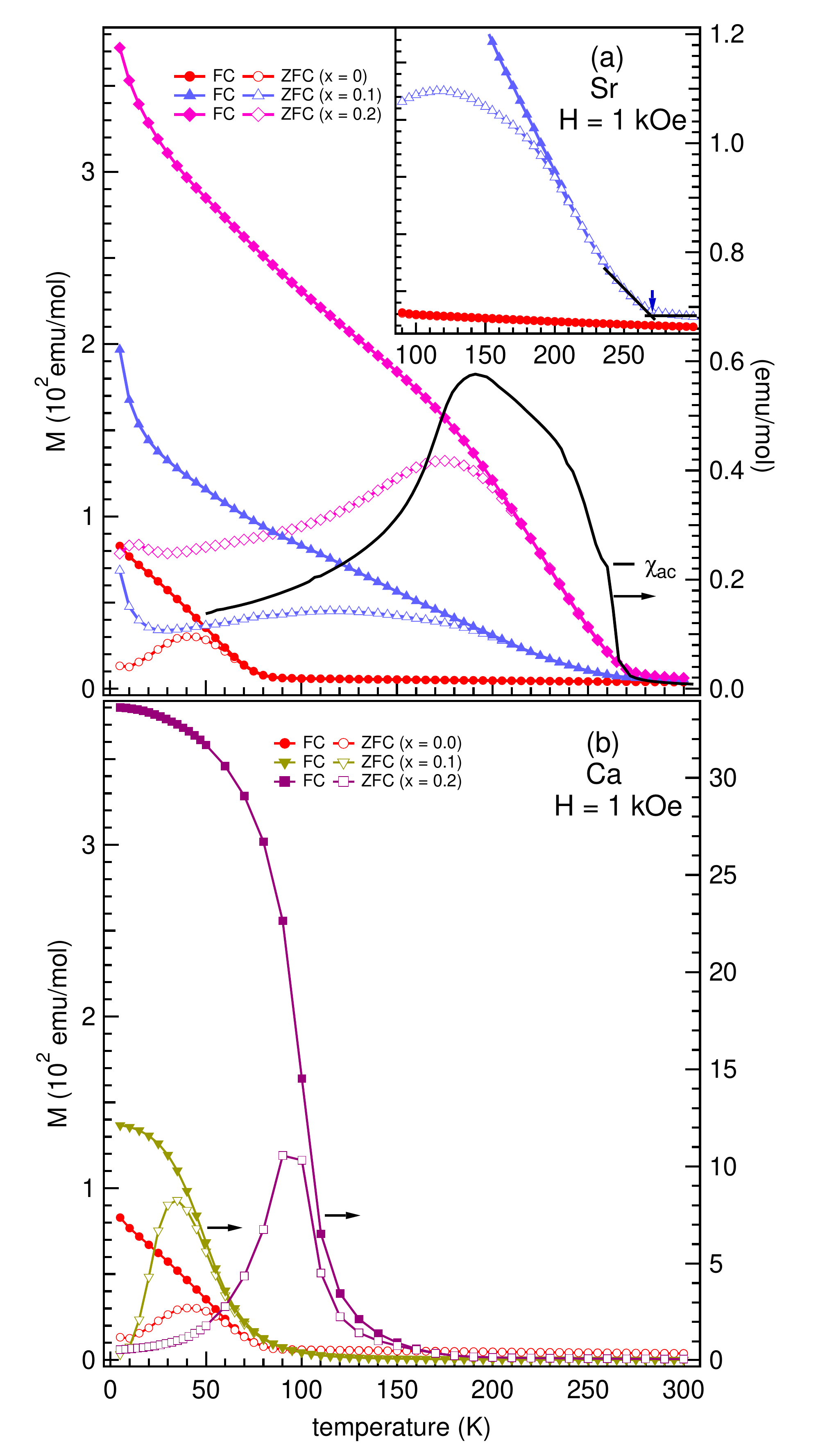}
\caption{(a) The temperature dependence of ZFC and FC magnetization data of La$_{(1-x)}$Sr$_x$CoO$_3$ nanoparticles and $\chi_{ac}$(T) of $x=$ 0.2 sample measured at 1~Oe and 1~Hz. (b) same as (a), but for La$_{(1-x)}$Ca$_x$CoO$_3$. These measurements are done at magnetic field H = 1~kOe.}
\label{fig4}
\end{figure}

\begin{figure}
\includegraphics[width=3.5in]{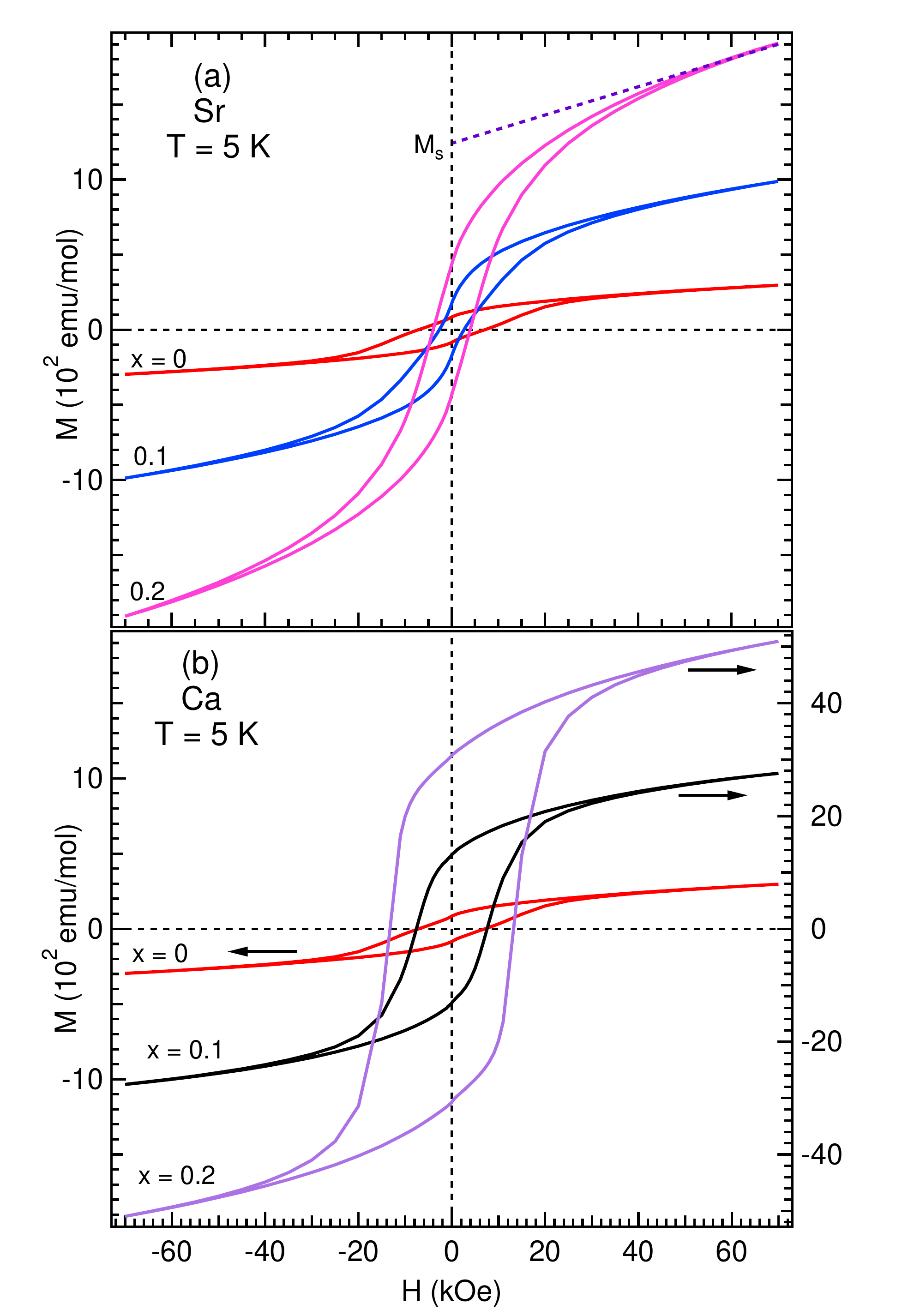}
\caption{(a) The field dependence of magnetization data of La$_{(1-x)}$Sr$_x$CoO$_3$ nanoparticles, measured at 5~K. (b) same as (a), but for La$_{(1-x)}$Ca$_x$CoO$_3$.}
\label{fig5}
\end{figure}

Note that the magnetic behavior of nanoparticle samples is dramatically different than the bulk. Further, we observed considerable changes in the magnetic properties of LaCoO$_3$ nanoparticles with the hole doping i.e. by substituting Sr/Ca at the La site. Figs.~4(a, b) show the temperature dependence of ZFC and FC magnetization data of La$_{(1-x)}$Sr$_x$CoO$_3$ and La$_{(1-x)}$Ca$_x$CoO$_3$ nanoparticles, respectively, measured at the magnetic field H = 1~kOe. Recently, we reported that by substituting small amount of Sr ($x =$ 0.05 sample) the T$_{\rm C}$ increases significantly i.e. from $\approx$80~K to $\approx$270~K \cite{Ravi_AIP17}. With increasing the Sr concentration, there is no significant change in the value of T$_{\rm C}$ up to $x =$ 0.2 sample, as determined by the onset of the magnetization and clearly visible/marked by an arrow in the inset of Fig.~4(a). The T$_{\rm C}$ value of each sample is further determined from the minimum in the derivative of the FC magnetization curves. Also, we determine the value of T$_{\rm C}$ more precisely by performing the temperature dependent ac susceptibility [$\chi_{ac}$(T)] measurements, see the Fig.~4(a), at applied ac magnetic filed (=1~Oe) and frequency (=1~Hz).  For Sr substituted nanoparticle samples, the temperature dependent magnetization in FM state is quite different from the $x =$ 0 sample as well as from the bulk samples \cite{AarboghPRB06, SamalJPCM11}. The FC curve shows continuous increase in magnetization below T$_{\rm C}$ while the appearance of a broad hump/peak at around 170~K in the ZFC magnetization of $x =$ 0.1 sample, which suggests a competition between different magnetic interactions. Interestingly, both ZFC and FC curves show a steep increase below about 20~K and reaches about 70 and 200~emu/mol (at 3~K) for $x =$ 0.1 sample. The magnetic behavior of $x =$ 0.2 sample is very different as the bifurcation appears at 200~K and a broad cusp is observed in the ZFC curve at about 180~K, compare to at 40~K in $x=$ 0 sample. This indicates that glasslike behavior is still present up to 20\% Sr substitution \cite{NamPRB99, SamalJAP09, SamalJAP12, MannaJAP13}. In Fig.~4(b), we show the magnetization curves of La$_{(1-x)}$Ca$_x$CoO$_3$. For $x=$ 0.1 sample, the T$_{\rm C}$ is fairly unaffected, but the magnetization increased significantly with slight shift in the peak position in ZFC curve (35~K) as well as change in the slope of FC curve at low temperature. For $x=$ 0.2 sample, T$_{\rm C}$ increases to about 150~K and the peak in the ZFC curve is observed at about 100~K \cite{BurleyPRB04}. The FC magnetization has increased to $\approx$ 390~emu/mol at 5~K, which is about 2.5 times higher than $x=$ 0.1 sample. It is evident that at the low temperatures, the behavior of the FC curves is like a ferromagnetic Brillouin function for both the Ca substituted samples \cite{Wu}. Above T$_{\rm C}$, the magnetization of both Sr and Ca substituted nanoparticle samples obey the Curie-Weiss law and the fitting to the linear part of an inverse susceptibility versus temperature curve allows the determination of an effective paramagnetic moment ($\mu_{\rm eff}$) and spin-state, discussed later. For all the samples, ZFC and FC curves diverge below T$_{\rm C}$, with the appearance of a peak in the ZFC data, which found to be much broad in Sr compare to the Ca samples. 

To further investigate the nature of the magnetic transition, in Fig.~5(a, b), we present the field dependence of isothermal magnetization $M$($H$) data measured at 5~K for La$_{1-x}$$A_{x}$CoO$_3$ ($A=$ Sr, Ca) nanoparticle samples. For $x =$ 0 sample, the observation of clear hysteresis loop confirms the existence of the ferromagnetic order at low temperatures below T$_{\rm C}$. The coercivity and spontaneous magnetization are found to be H$_{\rm C}=$ 7~kOe and M$_S$ = 180~emu/mol, respectively. In case of La$_{1-x}$Sr$_{x}$CoO$_3$, the hysteresis loop is much narrower with H$_{\rm C}=$ 2.7~kOe and the spontaneous magnetization increased (M$_S$ = 600~emu/mol) by about three times for $x =$ 0.1 sample. The coercivity and magnetization increases further to H$_{\rm C}=$ 4.4~kOe and M$_{\rm S}$ = 1050~emu/mol, respectively, for $x =$ 0.2 sample, see Fig.~5(a). On the other hand, in case of La$_{1-x}$Ca$_{x}$CoO$_3$, there is no change in coercivity, but the magnetization increases to 1900~emu/mol for $x =$ 0.1 sample. Furthermore, we observed H$_{\rm C}=$ 13.2~kOe and M$_{\rm S}=$ 3900~emu/mol for $x =$ 0.2 sample, which are dramatically larger than $x =$ 0 sample. These values are summarized in table~I for comparision. The M-H curve indicates easy magnetization and significant increase in the magnetic moment with hole doping. It should be noted that the field dependent magnetization measured at 5~K does not saturate even up to 70~kOe for both Sr and Ca substituted LaCoO$_3$ nanoparticles. This suggests that the magnetic ordering is more complex and not just a simple ferromagnetic order. Seo {\it et al.}, studied strained LaCoO$_3$ using {\it ab-initio} calculations and showed that due to strain there is competition between FM and AFM interactions in the system and proposed a canted spin structure \cite{SeoPRB12}. In the present case, due to nano-size effect and larger surface to volume ratio, strain develops which favors the spin canting due to coupling between FM and AFM. Moreover, we have estimated the spontaneous magnetization M$_{\rm S}$ by fitting the M--H curves with straight line above 50~kOe and summarized in the table~I. As mentioned above, in contrast to the Sr$^{2+}$ ion, the ionic radius of Ca$^{2+}$ ion is almost the same as the La$^{3+}$ ion. Therefore, the Ca substitution at La site introduces mainly holes into the system without creating a considerable crystal field distortion. Here, the observed enhancement in the ferromagnetic interactions with Sr/Ca substitution are due to double exchange mechanism between Co$^{3+}$ and Co$^{4+}$ ions.

\begin{figure}
\includegraphics[width=3.45in]{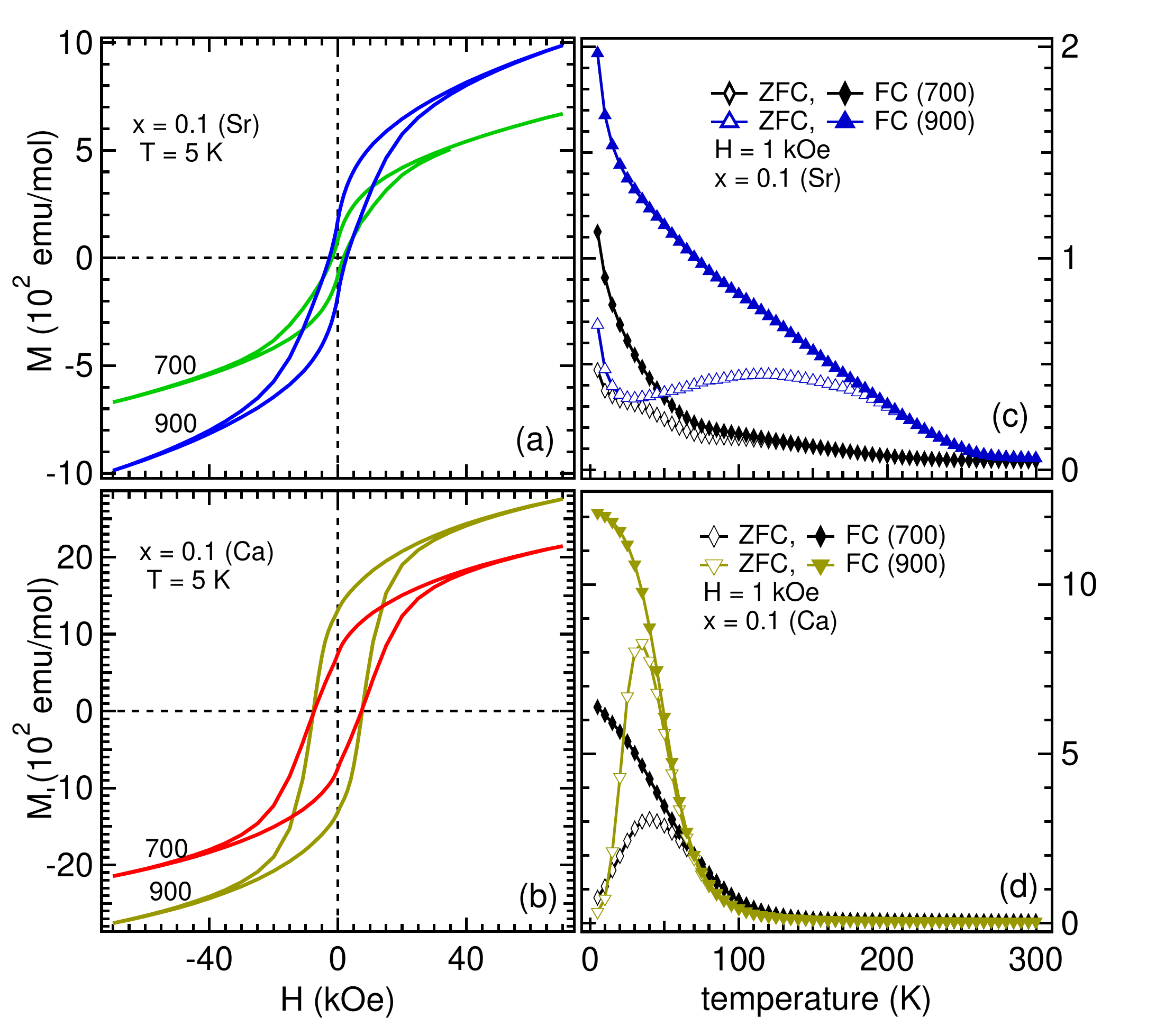}
\caption{The isothermal magnetization (a, b), measured at 5~K and magnetic susceptibility (c, d), measured at 1~kOe, of La$_{0.9}$A$_{0.1}$CoO$_3$ ($A=$ Sr, Ca) nanoparticles of two different sizes (annealed at 700$^o$C and 900$^o$C). }
\label{fig5}
\end{figure}

\begin{table*}
  \centering
  \caption{Experimentally obtained and calculated parameters, Curie constant C, spontaneous magnetization  $M_{\rm S}$,  remanence magnetization $M_{\rm r}$ (C in emu K mol$^{-1}$, $M_{\rm S}$ and $M_{\rm r}$ in emu mol$^{-1}$, $\mu$$_{eff}$ in $\mu_B$) of La$_{1-x}$$A_{x}$CoO$_3$ ($A=$ Sr, Ca) nanoparticles.}
  \label{tab:table2}
\vskip 0.3 cm
  \begin{tabular}{|c|c|c|c|c|c|c|c|c|c|c|c|}
  	\hline
   \textit{$x$ ($A$)} &  $M_{\rm S}$ & $M_r$ & $H_c$ & ${\rm C}$ & $\theta$$_{CW}$ &$\mu_{eff}$  &  S$_{avg}$ &  $\mu$$_{eff}$ & S$_{avg}$ & (1-$x$)$\times$Co$^{3+}$ & $x$$\times$Co$^{4+}$ \\
   \textit  &  &  & (kOe) &  &(K) &  (exp) & (exp)  &   (cal) & (cal) & IS:HS & LS \\
    \hline
    0.0 &  180 & 85 & 7.0 &1.86 & -197 &3.85 & 1.49 &  3.87 &1.5 & 50:50 &  0\\
    \hline
    0.05 (Sr) &  500& 65 & 0.6 &1.61 & -27&3.60 & 1.37 &  3.62 &1.38 & 57:43 & 100\\
    \hline
    0.10 (Sr)& 600 & 180 & 2.7 & 1.32 &61 &3.25 & 1.2 &  3.25 &1.2 & 72:28 & 100\\
    \hline
    0.20 (Sr) & 1050 & 375 & 4.4 &0.66 &193 & 2.30  & 0.75 & 2.6 &0.9 & 100:0 & 100\\
    \hline
    0.10 (Ca) &  1900& 1300 & 7.6 &1.40 &-15 & 3.34  & 1.24 & 3.37 & 1.26 & 65:35 & 100\\
    \hline
    0.20 (Ca) & 3900 & 3100 & 13.2 &1.36 & 97 & 3.30  & 1.22 & 3.21 &1.18 & 65:35 & 100\\
    \hline
  \end{tabular}
\end{table*} 

In order to study the effect of nano-size on magnetization, we prepared the samples by annealing at lower temperature (700$^o$C), which produce smaller particle size than the samples annealed at 900$^o$C. We compare the magnetization behavior of La$_{0.9}$A$_{0.1}$CoO$_3$ ($A=$ Sr, Ca) nanoparticles with different particle sizes, as shown in Fig.~6. At the smaller size, we observe significant decrease in spontaneous magnetization (M$_{\rm S}$ = 375 and 1500~emu/mol for Sr and Ca samples, respectively, annealed at 700$^o$C as compared to M$_{\rm S}$ = 600 and 1900~emu/mol for respective samples, annealed at 900$^o$C). However, there is no change in the coercivity, see Figs.~6(a, b). The temperature dependent magnetization of La$_{0.9}$Sr$_{0.1}$CoO$_3$ (700$^o$C sample) increases slowly below $\approx$ 200~K and the bifurcation appears below about 80~K where both FC and ZFC curves show increasing behavior at low temperatures. This is dramatically different from the sample annealed at 900$^o$C, as shown in Fig.~6(c). On the other hand for La$_{0.9}$Ca$_{0.1}$CoO$_3$, the magnetic susceptibility show similar temperature dependance for both the samples except that the moment of 700$^o$C sample below $\approx$80~K is much smaller than that of the 900$^o$C sample [Fig.~6(d)]. These results indicate the nano-size effect on the magnetic properties and reveal that the magnetization decreases with decreasing the particle size \cite{FitaJAP10}, which is in contrast with refs.~\cite{ZhouJPCC09, WangJAP10}.

\begin{figure}
\includegraphics[width=3.5in]{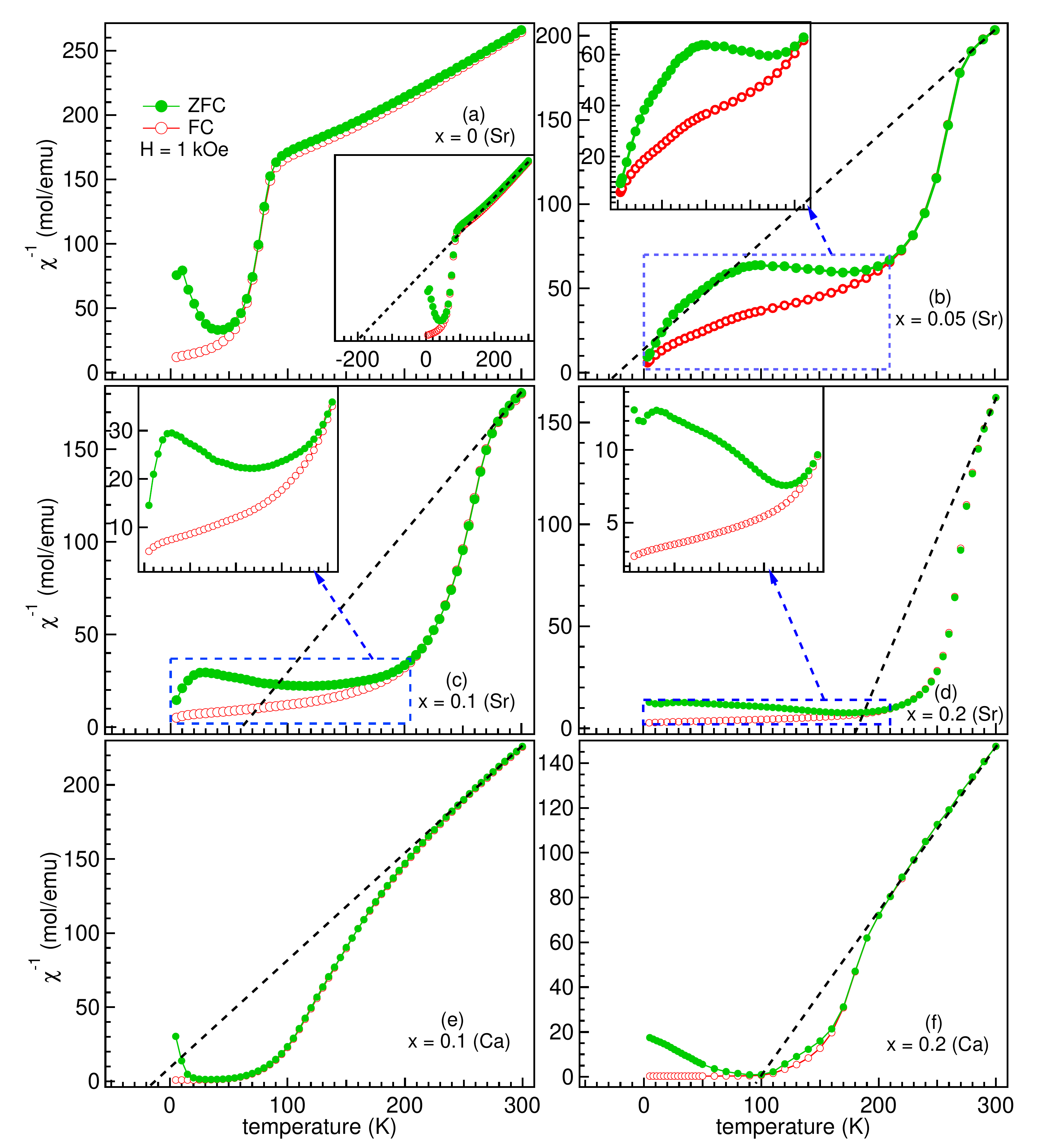}
\caption{ZFC and FC inverse magnetic susceptibility ($\chi^{-1}=$ H/M) data of La$_{1-x}$$A_{x}$CoO$_3$ ($A=$ Sr, Ca) nanoparticles.}
\label{fig6}
\end{figure}

In the following discussion, we focus on the temperature dependence of the magnetization and the spin-state transition in Sr and Ca substituted LaCoO$_3$ nanoparticle samples. Figs.~7(a--f) show the inverse magnetic susceptibility $\chi^{-1}$ (= H/M) of La$_{1-x}$$A_{x}$CoO$_3$ ($A=$ Sr, Ca) as a function of sample temperature measured at H = 1~kOe during ZFC and FC. The sharp decrease below T$_{\rm C}$ indicates the appearance of the ferromagnetic state in these nanoparticle samples, which is consistent with ref.~\onlinecite{ZhouPRB07}. The susceptibility at higher temperatures in the paramagnetic state obeys the Curie-Weiss (CW) law and fitted using $\chi^{-1} (T)=$ H/M = (T -- $\theta_{\rm CW}$)/C, where C is the Curie constant and is defined by N$\mu^{2}_{eff}$/3k$_B$. We fitted the high temperature linear part of $\chi^{-1}$ (see the dotted line in Fig.~7) and the experimentally obtained values of CW temperature ($\theta_{\rm CW}$), and C are given in table~I for all the samples. For $x =$ 0 sample, $\theta_{\rm CW}$ and C are found to be -197~K and 1.86~emu K mol$^{-1}$. The negative value of $\theta_{CW}$ suggests dominant AFM interactions in the nanoparticle samples and the increase in $\theta_{\rm CW}$ indicates that the FM interactions are enhanced with hole doping. The effective paramagnetic moment $\mu_{eff}$ can be obtained by $\mu_{eff}$=$\sqrt{8\rm C}$, as given in table~I. Also, the $\mu_{eff}$ can be written as $\mu_{eff}$=g$\sqrt{[J(J+1)]}$ $\mu_B$, where, g (= 2) and J are the Lande g factor, and total spin quantum number, respectively. Here, we consider the spin only value i.e. J = S$_{spin}$, which gives the value of average spin (S$_{avg}$). This means the slope of $\chi^{-1}$ vs temperature plot provides the spin value S$_{avg}$. The $\mu_{eff}$ and S$_{avg}$ values for $x =$ 0 sample (in table~I) are very close to the reported in ref.~\onlinecite{ZhouPRB07} and for strained films \cite{FuchsPRB07}.

Next we examine the possible spin state of Co ions in La$_{(1-x)}$$A_x$CoO$_3$ ($A=$ Sr, Ca) nanoparticles. Note that the radius of Co$^{3+}$ ions in HS ($\approx$0.61 $\rm {\AA}$) and IS ($\approx$0.56 $\rm {\AA}$) states is larger than the LS ($\approx$0.545 $\rm {\AA}$) state \cite{Shannon76}. Recently, we reported that by considering Co$^{3+}$ in both IS and HS states in 50 : 50 ratio,\cite{Ravi_AIP17} the $\mu_{eff}$(= 3.87 $\mu_B$) can be obtained close to the experimentally observed value for $x =$ 0 sample, see table~I. This is consistent with the fact that expansion in the unit cell volume in nanoparticles is due to the presence of Co$^{3+}$ in IS and HS states ($r_{\rm HS}$ $\textgreater$ $r_{\rm IS}$ $\textgreater$ $r_{\rm LS}$) compare to the bulk \cite{ZhouPRB07}. Photoemission studies of LaCoO$_3$ also reveal that there is decrease in the LS contribution at high temperatures \cite{BarmanPRB94, AbbatePRB93}. However, the crucial point here is to  find the spin state of Co$^{4+}$ ions, which are created into the Co$^{3+}$ matrix by Sr$^{2+}$ and Ca$^{2+}$ substitution at La$^{3+}$ site. The Co$^{4+}$ ions have 3d$^5$ electronic configuration and can present in LS; t$^5_{2g}$e$^0_g$ (S = $\frac{1}{2}$), IS; t$^4_{2g}$e$^1_g$ (S = $\frac{3}{2}$) or HS; t$^3_{2g}$e$^2_g$ (S = $\frac{5}{2}$) states. The radius of Co$^{4+}$ ions is about 0.54 $\rm {\AA}$ and 0.48 $\rm {\AA}$ in HS and LS states, respectively. The spin states of both Co$^{3+}$ and Co$^{4+}$ mainly depends on the competition between the crystal field splitting energy and the intra-atomic Hund's exchange interaction, which can be tuned by changing the temperature, applying magnetic field/pressure and by chemical substitution at La/Co sites. For different concentration level ($x$), we can write La$^{3+}_{1-x}$A$^{2+}_x$Co$^{3+}_{1-x}$Co$^{4+}_{x}$O$^{2-}_3$ ($A=$ Sr, Ca) and the total spin as S$_{spin}$ = (1--$x$)S$_3$ + $x$S$_4$, where S$_3$ and S$_4$ represent the spin of Co$^{3+}$ and Co$^{4+}$ ions, respectively. The similar approach has been used in the past \cite{FitaJAP10,RoyAPL08JAC11,Vinod14,Wu}. Note that the crystal field splitting decreases with Sr substitution due to the larger size of Sr ion, which is expected to enhance the IS state \cite{SatheJPCM96}. 
\begin{figure}
\includegraphics[width=3.5in]{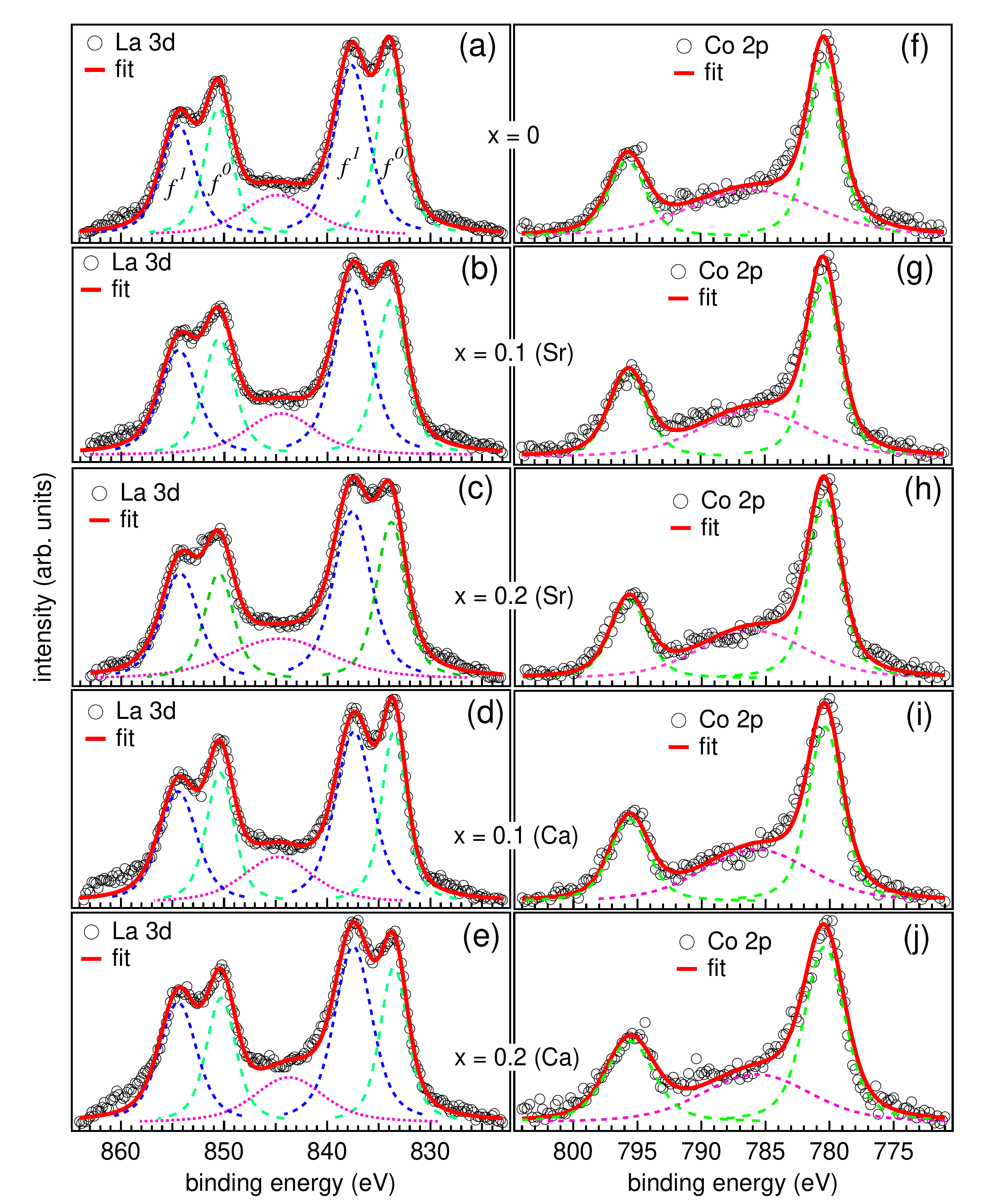}
\caption{(a) La 3$d$ and (b) Co 2$p$ core level spectra of La$_{(1-x)}$$A_x$CoO$_3$ ($A=$ Sr, Ca) nanoparticle samples.}
\label{fig7}
\end{figure}
On the other hand, Ca having similar ionic size as La may not affect the spin state significantly. In order to get the experimentally observed values of $\mu_{eff}$ and S$_{avg}$ for Sr substituted (up to $x=$ 0.2) samples, we tried different combinations of spin-state configurations for both Co$^{3+}$ and Co$^{4+}$ ions. On the basis of the calculations, we have summarized the percentage of the possible spin-states of both Co$^{3+}$ and Co$^{4+}$ ions in the table~I. For example, in $x=$ 0.2(Sr) sample we get 100\% Co$^{3+}$ ions in IS state and Co$^{4+}$ ions in LS state. Similarly, we found about 15\% increase in the IS of Co$^{3+}$ with Ca substitution, see table~I. Our calculation indicates that Sr/Ca substitution stabilize the IS state of Co$^{3+}$ and LS state of Co$^{4+}$ ions \cite{KrienerPRB04, BerggoldPRB05}. This is consistent with Ravindran {\it et al.}, as they showed that the HS state cannot be stabilized by temperature or hole doping as the HS state is significantly higher in energy than the LS or IS state \cite{RavindranJAP02}. They found the IS state to be metallic and the HS state is predicted to be half-metallic ferromagnet \cite{RavindranJAP02}. Using first-principles electronic structure calculations, it is reported that hole doping enhances Co-O hybridization, which favors the IS state and reduce the saturation magnetization compared to the ionic model \cite{RavindranJAP02}. Also, neutron diffraction studies show that the Sr substitution introduces Co$^{4+}$ ions in LS state that in turn stabilize near neighbor Co$^{3+}$ ions in IS state \cite{Caciuffo}. D. Phelan {\it et al.}, showed that a metallic ferromagnetic transition in La$_{1-x}$Sr$_{x}$CoO$_3$ results from the double-exchange coupling between IS state of Co$^{3+}$ and LS state of Co$^{4+}$ ions \cite{PhelanPRL06}. Our results reveal that the Sr/Ca substitution increases the population of IS (Co$^{3+}$) and LS (Co$^{4+}$) states, see table~I. The reduced JT splitting and the absence of LS state of Co$^{3+}$ are considered to be responsible for the emergence of ferromagnetism in hole doped LaCoO$_3$ nanoparticles \cite{WangJAP10}.

\begin{figure}
\includegraphics[width=3.4in]{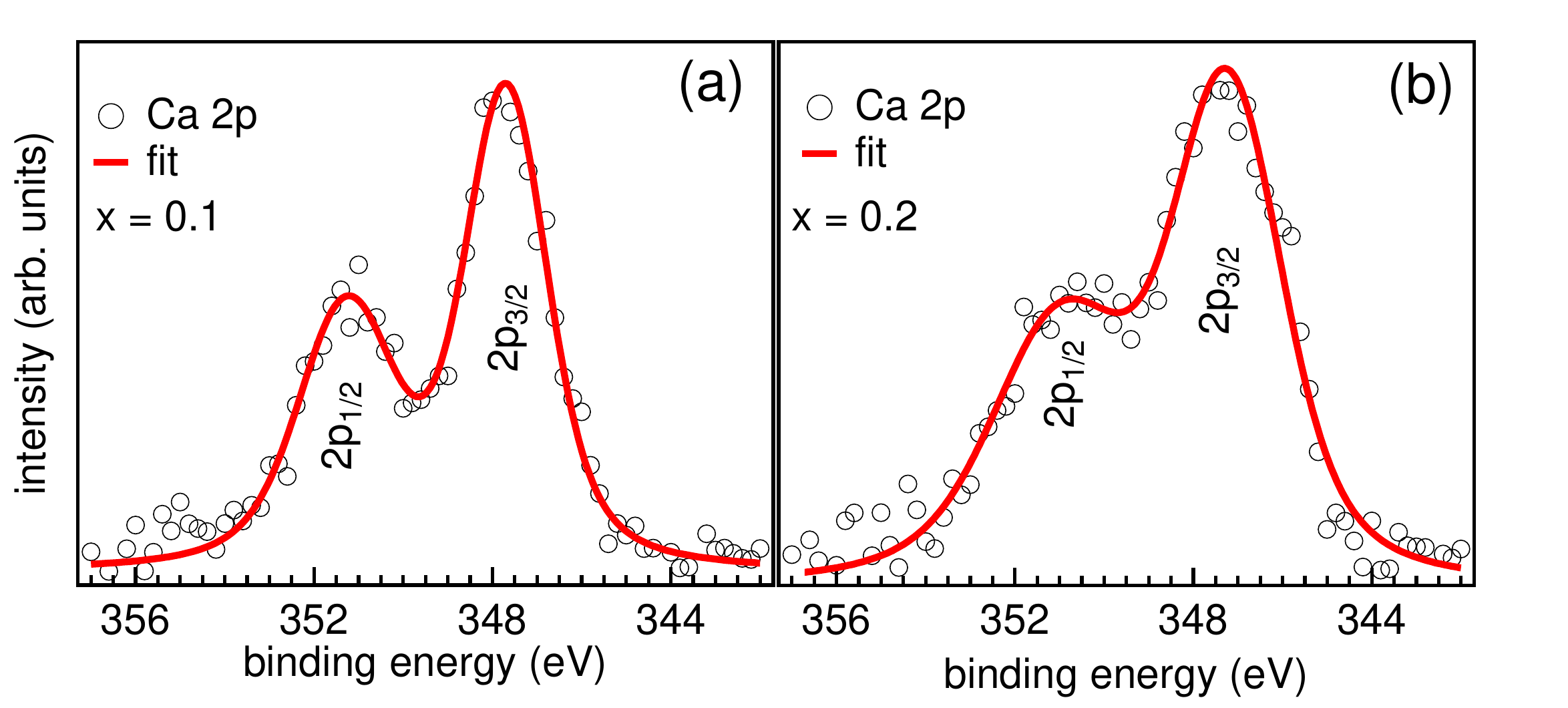}
\caption{Ca 2$p$ core level spectra of La$_{(1-x)}$Ca$_x$CoO$_3$ (a) $x=$0.1 and (b) $x=$0.2 nanoparticle samples.}
\label{fig8}
\end{figure}

In Figs.~8(a--e), we present the La 3$d$ and Co 2$p$ core-level spectra of La$_{(1-x)}$$A_x$CoO$_3$ ($A=$ Sr, Ca), measured at room temperature. The La 3$d$ core-level spectra show spin-orbit splitting components La 3$d_{5/2}$ and La 3$d_{3/2}$ at 834 and 850.7~eV binding energies (BEs), respectively. We observe a double-peak structure in each spin-orbit component, as denoted by $f^0$ and $f^1$ in Fig.~8(a). These peaks are corresponding to states with 3$d^9$4$f^0$ and 3$d^9$4$f^1$L configurations, where L denotes the hole in the O 2$p$ valence orbital (i. e. O 2$p^5$) \cite{VasquezPRB96}. The charge-transfer energy difference between these states is negligible, resulting in a strong interaction between them. The additional peaks ($f^1$) in both La 3$d_{5/2}$ and La 3$d_{3/2}$ are due to the transfer of an electron from the oxygen 2$p$ orbital to the empty La 4$f$ shell and attributed to final-state well screened satellites \cite{Lam80}. The spin orbit splitting of the La 3$d$ core-level and the energy separation between two peaks in each spin-orbit component are found to be around 16.7~eV (same as in La$_2$O$_3$, which suggests that the La ions are in the 3+ states) and 3.8~eV, respectively for $x=$0 sample. There is no significant change in the peak positions of spin orbit components with Sr/Ca concentration. However, with increasing Sr/Ca concentration we observe the decrease in the energy separation (3.8, 3.6, 3.4, 3.75, and 3.55~eV) and an increase in the intensity ratio (I$_{f^1/f^0}=$ 1.05, 1.15, 1.25, 1.45, and 1.5) for $x=$ 0, 0.1(Sr), 0.2(Sr), 0.1(Ca), and 0.2(Ca) samples, respectively, between the satellite peaks ($f^1$) with respect to the main core-level ($f^0$) peaks. These results indicate significant interactions between the La ions and their surrounding atoms \cite{Lam80}. Figs.~8(f--j) show the Co 2$p$ core-level spectra where the two main spin-orbit split peaks are observed at about 779~eV (2$p_{3/2}$) and 795~eV (2$p_{1/2}$) for $x=$ 0 sample, which indicates that Co is predominantly present in the Co$^{3+}$ state. The broad peak located around 788~eV is due to charge transfer satellite of the 2$p_{3/2}$ peak indicating correlation effects \cite{Lam80,Chainani92,Saitoh97}. There is no measurable shift (within the experimental accuracy) in the BE position of Co 2$p$ up to 20\% Sr/Ca substitution \cite{Chainani92,Saitoh97}. However, the presence of a shoulder on the high energy side, which implies the presence of Co$^{4+}$ states, requires further investigation with high energy resolution using synchrotron radiation facility, which is beyond the scope of this paper. We have also measured the Sr 3$d$ (not shown) and Ca 2$p$ [as shown in Figs.~9(a, b)] core level spectra \cite{Chainani92,VasquezPRB96}. The spin-orbit splitting components (2$p_{3/2}$ and 2$p_{1/2}$) in Ca 2$p$ appear at about 347.5~eV and 351~eV, respectively, which are higher than the reported in ref.~\cite{VasquezPRB96}. The determined surface compositions, by taking area under the core-level peaks and considering the respective photoionization cross-section,\cite{DhakaSS09}  are found to be close to the desired values.

\section{\noindent ~Conclusions}

In summary, we report the structural, magnetic and electronic properties of La$_{(1-x)}$$A_x$CoO$_3$ ($A=$ Sr, Ca; $x=$ 0 -- 0.2) nanoparticles using x-ray diffraction, magnetic susceptibility, isothermal magnetization, and x-ray photoelectron spectroscopy measurements. The Rietveld refinements of room temperature powder x-ray diffraction data confirm the crystalline nature and single phase of all the prepared samples. For $x=$ 0 sample, the magnetic measurements show a ferromagnetic transition at T$_C$$\approx$85 K, which shifted to higher temperatures with Sr/Ca substitution at La site. We also observe a significant increase in the spontaneous magnetic moment. Whereas, the coercive field H$_{\rm C}$ first decreases and then show increasing trend with Sr concentration. On the other hand, the behavior is dramatically different in Ca substituted samples. For example, the H$_{\rm C}$ values are 7, 4.4 and 13.2~kOe for $x=$ 0, 0.2 (Sr), and 0.2(Ca) samples. The increase in $\theta_{\rm CW}$ value indicates that the ferromagnetic interactions are enhanced with hole doping. The behavior of the FC magnetization is like a ferromagnetic Brillouin function at low temperatures for Ca substituted samples. Our study demonstrates that the Sr/Ca substitution increase the population of IS (Co$^{3+}$) and LS (Co$^{4+}$) states, which indicate that the double-exchange (Co$^{3+}$-- Co$^{4+}$) interactions are responsible for tuning the ferromagnetism. These results suggest an important role of hole carriers and nano-size effect in controlling the spin-state and magnetism by partial substitution of Sr/Ca at La site in LaCoO$_3$ nanoparticles.

\section{\noindent ~Acknowledgments}

RP, RS, and Priyanka acknowledge the MHRD, India for fellowship. Dr. V. K. Anand, and Dr. Milan Radovic are thanked for help in data analysis and useful discussions. Authors thank IIT Delhi for providing research facilities: XRD, SEM, TEM, XPS, SQUID, and ÒPPMS EVERCOOL-IIÓ. We also thank department of physics, IIT Delhi for support. RSD acknowledges the financial support from  SERB-DST through Early Career Research (ECR) Award (project reference no. ECR/2015/000159) and BRNS through ÒDAE Young Scientist Research AwardÓ project sanction No. 34/20/12/2015/BRNS.

\end{document}